\documentclass[twocolumn]{aastex62}

\newcommand{\bcep}{$\beta$~Cep }

\received{XXXX}
\revised{YYYY}
\accepted{ZZZZ}

\submitjournal{ApJL}

\shorttitle{Diverse variability of O and B stars with TESS}

\shortauthors{Pedersen et al.}

\begin{document}

\title{DIVERSE VARIABILITY OF O AND B STARS REVEALED FROM 2-MINUTE 
CADENCE LIGHT CURVES 
IN SECTORS 1 AND 2 OF THE TESS MISSION: SELECTION OF AN ASTEROSEISMIC 
SAMPLE} 

\correspondingauthor{May G. Pedersen}
\email{maygade.pedersen@kuleuven.be}


\author[0000-0002-7950-0061]{May G. Pedersen} \affiliation{Instituut voor
  Sterrenkunde, KU Leuven, Celestijnenlaan 200D, 3001 Leuven, Belgium}

\author[0000-0001-7444-5131]{Sowgata Chowdhury}
\affiliation{Nicolaus Copernicus Astronomical Center, Bartycka 18, 00-716
  Warszawa, Poland}

\author[0000-0002-3054-4135]{Cole Johnston} \affiliation{Instituut voor
  Sterrenkunde, KU Leuven, Celestijnenlaan 200D, 3001 Leuven, Belgium}

\author[0000-0001-7402-3852]{Dominic M. Bowman}
\affiliation{Instituut voor Sterrenkunde, KU Leuven, Celestijnenlaan 200D, 3001
  Leuven, Belgium}

\author[0000-0003-1822-7126]{Conny Aerts}
\affiliation{Instituut voor Sterrenkunde, KU Leuven, Celestijnenlaan 200D, 3001
  Leuven, Belgium}
\affiliation{Department of Astrophysics, IMAPP, Radboud University Nijmegen, PO Box 9010, NL-6500 GL Nijmegen, the Netherlands}

\author[0000-0001-7756-1568]{Gerald Handler} \affiliation{Nicolaus Copernicus Astronomical Center,
  Bartycka 18, 00-716 Warszawa, Poland}
  
\author[0000-0001-5419-2042]{Peter De Cat}
\affiliation{Royal Observatory of Belgium, Ringlaan 3, 1180 Brussel, Belgium}

\author[0000-0003-1978-9809]{Coralie Neiner}
\affiliation{LESIA, Paris Observatory, PSL University,  CNRS, Sorbonne Universit{\'e}, Univ. Paris Diderot, Sorbonne Paris Cit{\'e}, 5 place Jules Janssen, 92195 Meudon, France}

\author[0000-0003-4062-0776]{Alexandre David-Uraz}
\affiliation{Department of Physics and Astronomy, University of Delaware,
  Newark, DE 19716, USA}
  
\author[0000-0002-1988-143X]{Derek Buzasi}
\affiliation{Dept. of Chemistry \& Physics, Florida Gulf Coast University, 10501
  FGCU Blvd. S., Fort Myers, FL 33965 USA}

\author[0000-0003-0842-2374]{Andrew Tkachenko}
\affiliation{Instituut voor Sterrenkunde, KU Leuven, Celestijnenlaan 200D, 3001
  Leuven, Belgium}

\author[0000-0003-1168-3524]{Sergio Sim{\'o}n-D{\'i}az}
\affiliation{(Instituto de Astrof{\'i}sica de Canarias, 38200 La Laguna, Tenerife, Spain}

\author[0000-0003-4372-0588]{Ehsan Moravveji}
\affiliation{Instituut voor Sterrenkunde, KU Leuven, Celestijnenlaan 200D, 3001 Leuven, Belgium}

\author[0000-0002-3522-5846]{James Sikora}
\affiliation{Department of Physics, Engineering Physics and Astronomy, Queen’s University, Kingston, ON K7L 3N6 Canada}

\author[0000-0003-0238-8435]{Giovanni M. Mirouh}
\affiliation{Astrophysics Research Group, Faculty of Engineering and Physical Sciences, University of Surrey, Guildford GU2 7XH, UK}

\author[0000-0003-4233-3105]{Catherine C. Lovekin}
\affiliation{Physics Department, Mount Allison University, Sackville, NB, E4L 1C6, Canada}

\author[0000-0002-8171-8596]{Matteo Cantiello}
\affiliation{Center for Computational Astrophysics, Flatiron Institute, 162 5th Avenue, New York, NY 10010, USA}
\affiliation{Department of Astrophysical Sciences, Princeton University, Princeton, NJ 08544, USA}

\author[0000-0001-9704-6408]{Jadwiga Daszy{\'n}ska-Daszkiewicz}
\affiliation{Instytut Astronomiczny, Uniwersytet Wroc{\l}awski, Kopernika 11, 51-611 Wroc{\l}aw, Poland}

\author[0000-0003-2488-6726]{Andrzej Pigulski}
\affiliation{Instytut Astronomiczny, Uniwersytet Wroc{\l}awski, Kopernika 11, 51-611 Wroc{\l}aw, Poland}

\author[0000-0001-6763-6562]{Roland K. Vanderspek}
\affiliation{Department of Physics, and Kavli Institute for Astrophysics and Space Research, Massachusetts Institute of Technology, Cambridge, MA 02139, USA}

\author{George R. Ricker}
\affiliation{Department of Physics, and Kavli Institute for Astrophysics and Space Research, Massachusetts Institute of Technology, Cambridge, MA 02139, USA}


\begin{abstract}
  Uncertainties in stellar structure and evolution theory are largest for stars undergoing core convection on the main sequence. A powerful way to calibrate the free parameters used in the theory of stellar interiors is asteroseismology, which provides direct measurements of angular momentum and element transport. We report the detection and classification of new variable O and B stars using high-precision short-cadence (2-min) photometric observations assembled by the {\it Transiting Exoplanet Survey Satellite} (TESS). In our sample of 154 O and B stars,  we detect a high percentage (90\%) of variability. Among these we find 23 multiperiodic pulsators, 6 eclipsing binaries, 21 rotational variables, and 25 stars with stochastic low-frequency variability. Several additional variables overlap between these categories. Our study of O and B stars not only demonstrates the high data quality achieved by TESS for optimal studies of the variability of the most massive stars in the Universe, but also represents the first step towards the selection and composition of a large sample of O and B pulsators with high potential for joint asteroseismic and spectroscopic modeling of their interior structure with unprecedented precision.
\end{abstract}



\section{Introduction}
\label{section: intro}

The variability of stars born with a mass $M\geq 3\,$M$_\odot$ is
diverse in terms of periodicity (minutes to centuries) and amplitude ($\mu$mag
to mag, see e.g. \citealt{Aerts2010}). Here, we are concerned with dwarfs, giants, and supergiants, of spectral type O or B. Such
stars have a high binarity rate, a
phenomenon that cannot be ignored when testing stellar evolution
theory \citep{Sana2014,Almeida2017,Schneider2018}.  Throughout
their life, these stars are also subject to a strong variable
radiation-driven wind \citep[e.g.][]{Lucy1970,Castor1975,Owocki1984,Krticka2018}.

Compared to other classes of variables, O stars have
hardly been monitored with high-precision long-duration 
space photometry -- see Table\,3 in
\citet{Buysschaert2015} for a summary of space photometric time-series studies
of O stars, and more recent studies of their descendants  including
\citet{Pablo2017,Johnston2017,Buysschaert2017a,Aerts2018a,SSD2018,Ramia2018a,Ramia2018b}. Aside
from single and multiperiodic pulsational, binary and rotational variability, these 
studies also revealed stochastic low-frequency variability. This observed phenomenon was
interpreted in terms of internal gravity waves (IGWs) by
\citet{AertsRogers2015}. The observational signatures of IGWs have meanwhile been
investigated systematically from CoRoT and K2 data by \citet{Bowman2018} and
Bowman et al.\ (submitted), respectively. They offer an entirely new way of asteroseismic
investigation by bridging 3D hydrodynamical simulations of stochastically
excited waves and 1D theory of stellar interiors \citep[][Fig.\,1]{Aerts2019}. 

Classical gravity-mode asteroseismology, i.e., forward modeling of the
frequencies of coherent identified gravity modes \citep[see][for the methodology]{Aerts2018b}, has so far been limited to some 40 stars of spectral
type B or F, covering the mass range $M \in [1.3, 8]$~M$_{\odot}$ and rotation rates
from almost zero up to 80\% of critical breakup.  This revealed the capacity of
high-precision ($\sim 10\%$) mass estimation of single stars \citep[see][for
B stars and Mombarg et al.\ (submitted) for F
stars]{Moravveji2015,Moravveji2016,Szewczuk2018} and of binaries
\citep{Kallinger2017,Johnston2019}.  So far, these asteroseismic studies using space photometry
revealed almost rigid rotation for the stars whose
near-core ($\Omega_{\rm core}$) and envelope ($\Omega_{\rm env}$) rotation could
be estimated, following the few earlier results of
$\Omega_{\rm core}/\Omega_{\rm env}\in [1,5]$ from ground-based asteroseismology
of early-B stars \citep[see][for a summary]{Aerts2015}. These asteroseismic findings 
challenge current angular momentum transport
theories across the entire mass range \citep{Aerts2019}.

This {\it Letter\/} introduces our dedicated study to investigate single and
binary O and B stars with the NASA \emph{Transiting Exoplanet Survey Satellite} (TESS)  mission \citep{Ricker2015}. In order to perform 
asteroseismology, one first needs to find suitable multiperiodic O and B pulsators. We achieve this by classifying the variability of 154 O and B stars observed in and proposed for short cadence mode by the \textit{TESS Asteroseismic Science Consortium} (TASC). We present our classification and selection
strategy for the 2-min cadence light curves obtained in Sectors\,1 and 2 of the
TESS mission. 


\section{Method}
\label{section: method}


	\subsection{TESS light curve extraction}
	\label{subsection: data}

        To study the variability of massive stars, we analyse a sample of 154 O and B stars
        observed by TESS with 2-min cadence, of which 40 are located in the Large Magellanic Cloud (LMC). 
        These 154 stars were identified as O and B stars based on the spectral types provided in \texttt{SIMBAD}\footnote{\url{http://simbad.u-strasbg.fr/simbad/}}.
        The data treated here were obtained by TESS in Sectors\,1 (July 25 -- August
        22, 2018) and 2 (August 23 -- September 20, 2018) and are publicly available via
        the Mikulski Archive for Space Telescopes
        (MAST)\footnote{\url{http://archive.stsci.edu/tess/all_products.html}}. The
        extracted time series are in the format of reduced Barycentric Julian
        Date (BJD -- 2457000) and stellar magnitudes, where the latter have been
        adjusted to show variability around zero by subtracting the mean.
        Where necessary, we performed detrending of long-term instrumental effects by means 
        of subtracting a linear or low-order polynomial fit.


	\subsection{Classification procedure}
	\label{subsection: freq analysis}

        We calculated amplitude spectra of all light curves using Discrete
        Fourier Transforms (DFT) following the method by \citet{Kurtz1985}.  We
        used an oversampling factor of ten to ensure that all frequency peaks
        in the DFT are adequately sampled. The 2-min cadence of TESS
         results in a Nyquist frequency of 360~d$^{-1}$, which is
        sufficiently high to avoid a bias when extracting significant
        frequencies using iterative prewhitening. Amplitude suppression of
        variability in time series is negligible within the frequency range of interest  for such high
        sampling \citep{Bowman2017}. 
Based on visual inspection of the light curves and amplitude spectra by several of the authors independently, we provide the 
variability classification of all 154 O and B stars in Table~\ref{table: stars} in Appendix~\ref{appendix:
  table}. We also report the number of available \'echelle spectra in the public ESO archive\footnote{\url{http://archive.eso.org/eso/eso_archive_main.html}} for each of the stars and the instrument used for the observations. This information is relevant for future studies of these 154 O and B stars. Figures of the light curves and amplitude spectra are made available
electronically in Appendix~\ref{appendix: figures}.


	\subsection{Gaia color-magnitude diagram}
	\label{subsection: Gaia CMD}
	
        Asteroseismic modeling of stars with coherent oscillation modes can be
        optimally performed if at least one additional global stellar parameter (aside from the identified oscillation frequencies) can be measured with high precision. The availability of such an independent
        measurement helps to break degeneracies 
        among the global and local stellar model parameters to be estimated
        \citep[e.g.,][Figs\,5 and B.1]{Moravveji2015}.  A model-independent mass
        from binarity \citep{Johnston2019} or a high-precision (10\%)
        spectroscopic estimate of the effective temperature (Mombarg et al.,
        submitted) have been used to break degeneracies. Another useful quantity
        is a star's luminosity (Pedersen et al., in preparation).

\begin{figure*}
\center
\includegraphics[width=0.75\textwidth]{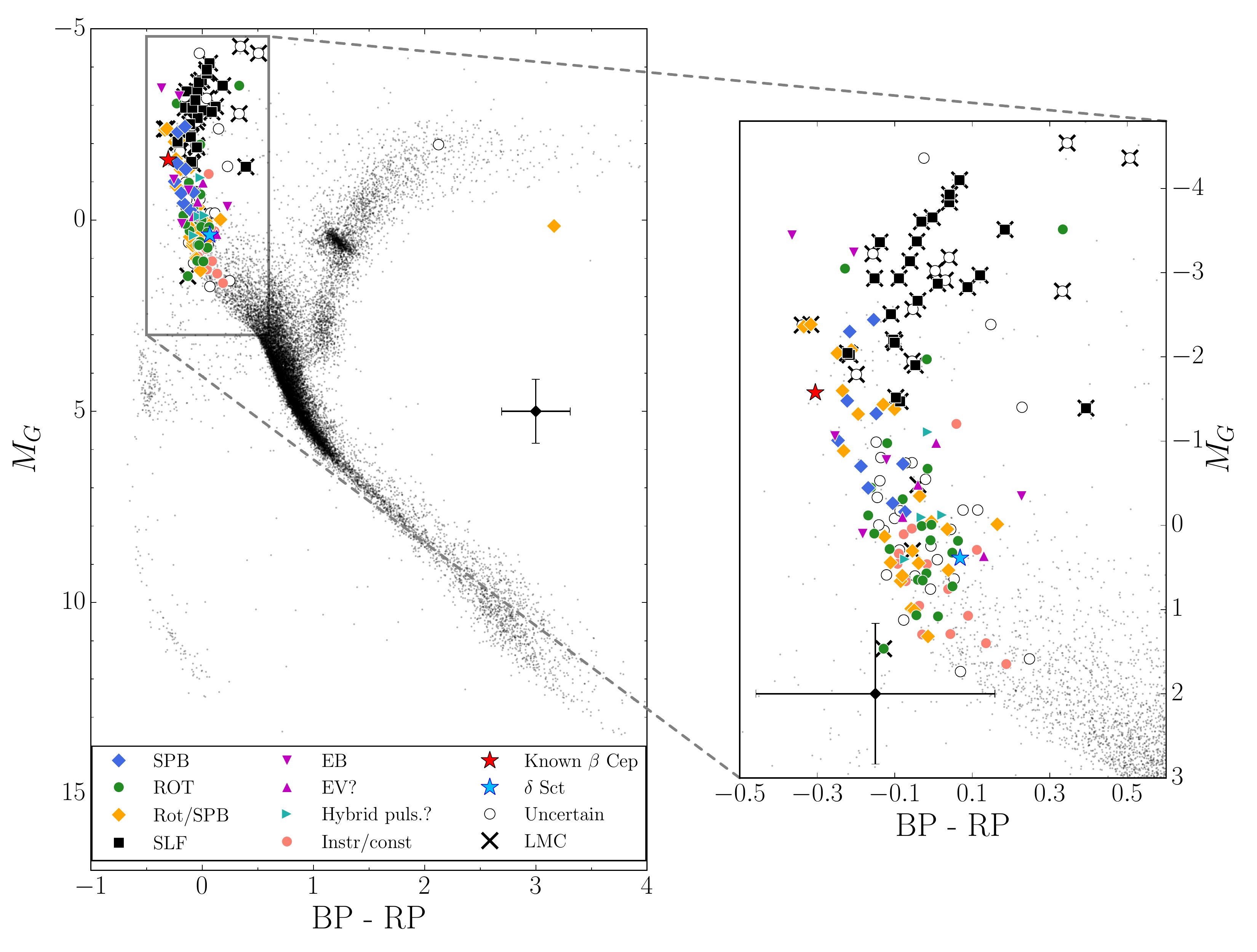}
\caption{Classification results for O and B stars placed in a Gaia DR2
  CMD. Black dots are the entire TESS sector 1 and 2 short-cadence stars also
  observed by Gaia, and open black circles corresponds to uncertain
  classification. The labels in the legend correspond to the following types of
  variability: SLF = Stochastic low-frequency signal, Instr/const = instrumental
  or constant, SPB = slowly pulsating B star, ROT = rotational modulation,
  ROT/SPB = rotational modulation and/or SPB, Hybrid puls.? = both p- and
  g-modes, EB = Eclipsing Binary, EV? = Ellipsodal variable or rotational
  modulation, $\delta$ Sct = $\delta$ Scuti star. The error bar shows a typical
  $2\sigma$-error on the position in the CMD, shown in both panels. LMC members
  are indicated by a {\large\bf{$\times$}} behind their variability symbol.}
\label{fig:CMDoverview}
\end{figure*}

	With this in mind, and to check the \texttt{SIMBAD} spectral types that went into
        the selection of our sample, we used Gaia DR2 photometry \citep{GaiaDR2}
        to place all sample stars in a color-magnitude diagram (CMD,
        Fig.\,\ref{fig:CMDoverview}). Each star was color-coded by its
        dominant variability type. The apparent Gaia G-band magnitudes were
        converted to absolute magnitudes $M_G$ using the Gaia DR2 distances from
        \citet{BailerJones2018}. The colors were derived from the apparent Gaia BP and RP band magnitudes. All other stars observed in 2-min cadence by
        TESS in Sectors\,1 and 2 with Gaia DR2 data available (20883 out of the 24816 TESS targets) were included as black dots. Using the same approach as in \citet{GaiaDR2variability}, the position of the stars in Fig. \ref{fig:CMDoverview} have not been corrected for reddening or extinction, but we show a typical $2\sigma$ error bar for the positioning of the stars in the CMD. Two stars (TIC~197641601 = HD~207971 and TIC~354671857 = HD~14228) are outliers in the CMD (at BP-RP $ > 2$). 
        Although these stars are B stars, they are very bright (V = 3.01 and 3.57, respectively), which explains the discrepant Gaia photometry.

        The 151 out of 154 stars in our sample observed with Gaia shown in
        Fig.\,\ref{fig:CMDoverview} constitute the first sample of
        variable O and B stars monitored at high cadence in high-precision space
        photometry, after the K2 sample monitored at 30-min
        cadence in Bowman et al.\ (submitted). These two samples, along with
        future ones assembled by the TESS mission, will reveal numerous O and B stars suitable for asteroseismic modeling. Such modeling
        requires a frequency precision better than 0.001\,d$^{-1}$ and pulsation
        mode identification for tens of modes \citep{Aerts2018b}. In this way, we
        will extend the large {\it Kepler\/} samples of
        low-mass and intermediate-mass pulsators with estimation of their
        interior rotation profile discussed in \citet{Aerts2019} to higher masses.  The interior
        rotation and chemical mixing profiles, will hence be calibrated asteroseismically for large
        samples of massive stars that will eventually explode as supernovae.

\section{Results}
\label{section: results}

It is not possible to discuss each star in the sample
individually. Here we focus on the binarity and pulsational properties, and briefly discuss the detection of rotational and/or magnetic properties of the sample. 
Out of 154 stars, 41 show
clear variability in their light curves but their nature could not yet be uniquely
identified. These stars are marked by open black circles in Fig.\ref{fig:CMDoverview}. 


	\subsection{Eclipsing binaries}
	\label{subsection: EB}

        Eclipsing binaries (EBs) hold the potential to provide model-independent
        distance, radius, and mass estimates and are crucial
        calibrators for stellar evolution theory \citep[][for a
        review]{Torres2010}. Unfortunately, the number of O and B stars in EBs
        observed with high-precision space photometry remains small compared to
        the thousands of binaries with low-mass components \citep{Kirk2016}.  

        Six of the stars in our sample were already known as EBs and we
        confirm this classification using the TESS data: HD\,6882 ($\zeta$ Phe),
        HD\,61644 (V455 Car), HD\,224113 (AL Scl), HD\,42933 ($\delta$ Pic), HD\,31407 (AN Dor), and HD\,46792 (AE Pic).
        Their light curves in Appendix~\ref{appendix: figures} are of
        unprecedented quality and future modeling will improve their mass
        determinations. One of these stars (HD\,42933) is known to have \bcep type pulsations detected from BRITE photometry \citep{Pigulski2017}. We confirm the variability. In addition, three objects (HD\,224990, HD\,269382 and HD\,53921) were
        known spectroscopic binaries, while ten more (HD\,269676,
        HD\,68520, HD\,19400, HD\,53921, HD\,2884, HD\,46860, HD\,208433,
        HD\,37854, HD\,209014 and CPD-60 944) were listed as double or multiple stars in
        \texttt{SIMBAD}; the TESS light curves of these 13 non-eclipsing binaries did not
        show traces of the binary nature. 
        The only exception is HD 208433, which shows a single 
        transit. In addition, we find four showing 
        either ellipsoidal variation or rotational modulation: HD\,268798, 
        HD\,222847, HD\,20784 and HD\,37935. None of these
        four stars have previously been identified as binaries, but HD37935 is a known Be star.

		HD\,53921 is a known spectroscopic magnetic binary, which was identified
        as a Slowly Pulsating B (SPB) star by \citet{DeCatAerts2002}. From
        the high-quality TESS light curves we find that the dominant frequency of 0.6054\,d$^{-1}$ shows a second harmonic. Based on this as well as the morphology of the light curve, we deduce that this frequency is caused by rotational modulation. 
        This example illustrates how difficult g-mode asteroseismology from ground-based observations can be. This and other known magnetic stars in the sample are discussed in detail in David-Uraz et al. (in prep.).
        
\begin{figure}
\includegraphics[width=\columnwidth]{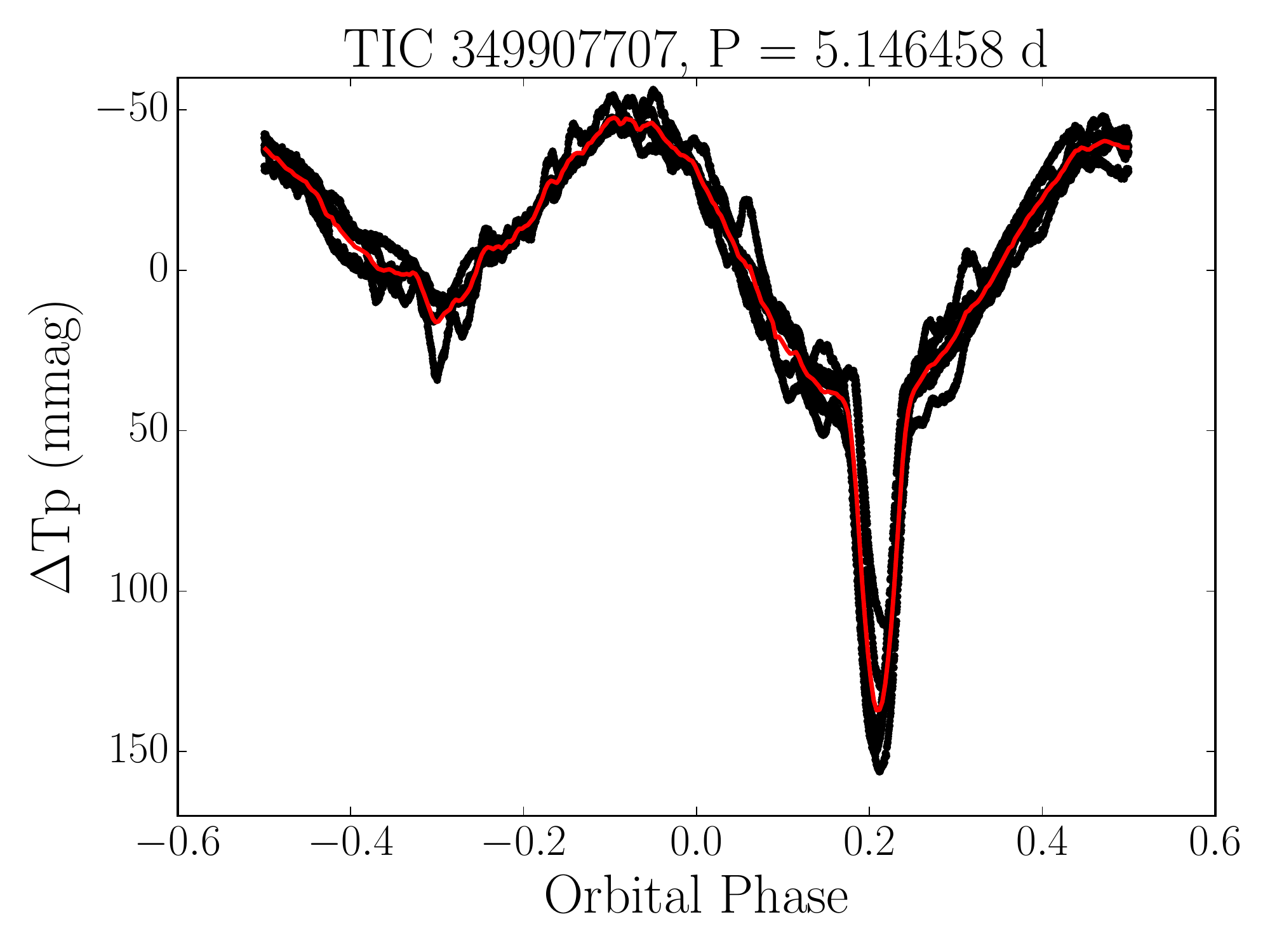}
\caption{Phase folded light curve (black) of the known eclipsing binary HD\,61644,
showing the signature of pulsations. The red line shows the binned phase curve.
} 
\label{Fig-EB} 
\end{figure}

In total, eight of these 19 binaries reveal
pulsations.
We show the phased light curve for a newly discovered binary pulsator (HD\,61644=TIC\,349907707) in Fig.\,\ref{Fig-EB}. This result illustrates
the promise of TESS to provide numerous pulsating O and B binaries suitable to
calibrate stellar evolution models from binary asteroseismology \citep{Johnston2019}.

	\subsection{Pulsating stars}
	\label{subsection: pulsators}
	
        Coherent non-radial oscillation modes in O and B stars come in two main
        flavors: pressure modes with periods of a few hours (\bcep stars,
        spectral types from O9 to B3) and gravity modes with periods of order a few
        days (SPB stars, spectral types from late O to B9). For a
        discussion on their early discoveries and pulsation properties, we refer
        to \citet[][Chapter 2]{Aerts2010}. The space-based photometry
      assembled with MOST, CoRoT, {\it Kepler\/}, K2, and BRITE revealed that
      many O and B pulsators are {\it hybrids}, i.e., pulsators
      with both types of modes simultaneously.
      Such hybrid pulsators have proven to be a powerful tool for constraining opacities in the partial ionization layers responsible for the mode excitation. As an example, the detailed seismic modeling of $\nu$ Eri performed by \citet{DaszynskaDaszkiewicz2017} revealed that a factor three increase in the opacity at $\log T = 5.46$ was needed to excite the g-modes in this star.
      
\begin{figure}
\includegraphics[width=\columnwidth]{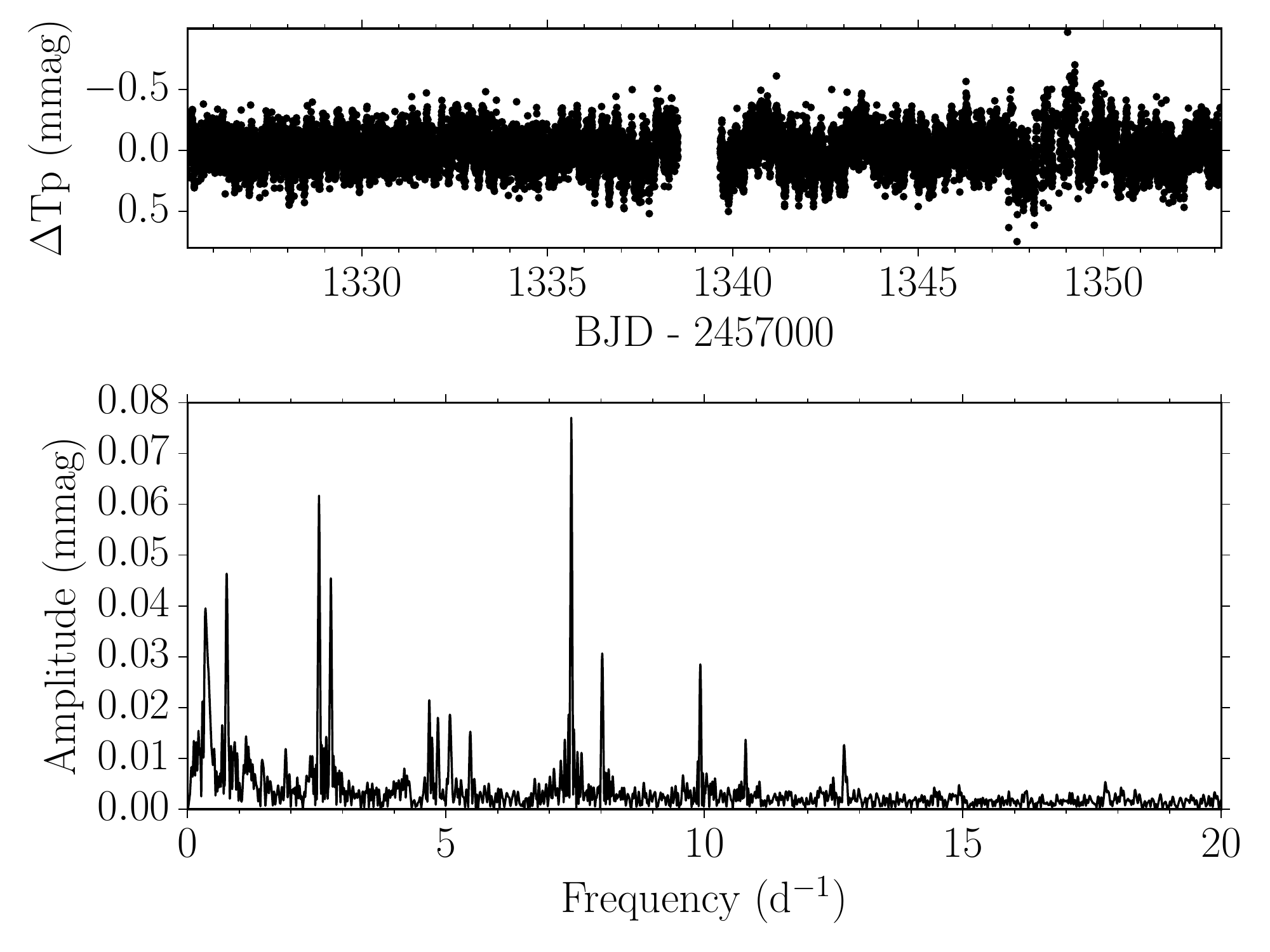}
\caption{Example of a newly-discovered hybrid pulsator showing both SPB and \bcep type pulsations (TIC~469906369 = HD\,212581). The TESS light curve and amplitude spectrum are shown in the top and bottom panels, respectively.} 
\label{Fig-Hybrid} 
\end{figure}

      In total, we find 14 new O and B pulsators, among which 10
      have gravity modes and four are potential hybrids. We show the light curve and DFT for one of
      the newly discovered hybrid stars in Fig.\,\ref{Fig-Hybrid}. One star in the sample, HN Aqr, is a known \bcep star and is discussed in Handler et al. (submitted). Future continued TESS and/or
      spectroscopic monitoring will be needed to assess the asteroseismic
      potential of these 14 pulsators in terms of frequency precision and mode identification. 

	\subsection{Rotational variability}
	\label{subsection: rot}
	
        We classified 21 of the stars in our sample as rotational variables, one of which is 
        labeled as a Wolf Rayet star in \texttt{SIMBAD} (HD\,269582).
        Among these are three stars previously known to be magnetic: HD\,223640 
        \citep{Bychkov2005,Sikora2018} and HD\,53921 \citep{Hubrig2006,Bagnulo2015}, as well as HD\,65987
        \citep{Bagnulo2015}. Rotational modulation is usually interpreted as due
        to temperature and/or chemical spots on the stellar surface caused by large-scale magnetic fields.  We refer to the accompanying papers by David-Uraz
        et al.\ (in prep.) and Balona et al.\ (submitted) for an extensive
        discussion of the rotational modulation and magnetic properties of the
        sample.

The star HD\,10144 (aka Achernar and $\alpha$\,Eri)
        is a Be star rotating at 95\% of critical breakup, whose
        stellar disk has been imaged by interferometry \citep{Achernar2017}.
We find its dominant frequency to be 0.729\,d$^{-1}$.  
Also HD\,214748, HD\,209522, HD\,33599, HD\,19818, HD\,221507, HD\,46860, HD\,37974, HD\,53048, CD-56 152, HD\,209014, HD\,37935, HD\,68423, HD\,224686 and HD\,66194 were known to be Be stars prior to our study. While HD\,221507 is known as a Be star, recent studies have shown that it lacks emission and has a low $v\sin i$ \citep[e.g.][]{Arcos2017}. 

		In addition to the stars classified as rotational variables, 
        we found 25 stars which show rotational modulation and/or SPB type variability. 
        The simultaneous occurrence of rotation and
        pulsation frequencies in space photometry is not unusual, but there is
        only one case so far for which combined spectropolarimetric and
        asteroseismic modeling has been achieved: the hybrid magnetic
        $\beta\,$Cep/SPB star HD\,43317 monitored by CoRoT
        \citep{Buysschaert2017b,Buysschaert2018}. Our results are 
        encouraging to expand the domain of magneto-asteroseismology. 

\subsection{Stochastic, low-frequency signatures}

Stochastic, low-frequency (SLF) variability has been shown to be a common phenomenon in blue supergiants in both space-based photometry \citep{Bowman2018} and spectroscopy \citep{SSD2018}. Several theories may explain this variability, such as sub-surface convection, dynamical stellar winds and IGWs excited at the convective core boundary \citep[see][for a detailed discussion]{Bowman2018}. In our sample of 154 O and B stars, we find 25 stars with SLF, all of which are classified as blue supergiants based on the spectral types from \texttt{SIMBAD} and occuring near the terminal age main sequence (TAMS). This is in agreement with the conclusions by Bowman et al. (submitted) from K2 photometry of such stars, and opens up an entirely new avenue of forward modeling of evolved OB stars at and beyond the TAMS to tune their angular momentum and element transport observationally.

\subsection{Other types of variability}

We found several stars with ``outbursts'' in their light curves (see Table~\ref{table: stars} and
Appendix~\ref{appendix: figures}). Such outbursts may be connected with episodic
mass loss, such as in Be stars, or flaring due to magnetic activity in low-mass
stars. In some cases, the detected outbursts occur in various of the light curves at the same time and for the same duration. This signature is instrumental. 
For the remainder of the targets, spectroscopic follow-up and a study of the pixel data is required to confirm that the outbursts are physical, rather than instrumental artefacts (\citealt[][Balona et al. submitted]{Pedersen2017}).


\section{Discussion and Conclusions}
\label{section: discussion}

We presented variability classification for 154 stars with spectral type O or B
that were monitored in short-cadence (2-min) by the TESS mission in its
Sectors\,1 and 2.  We found 138 to be variable at the TESS detection threshold.
This is a high percentage of variability (90\%) given that the time base of each
sector is only 27\,d, preventing longer-period variables from being
discovered. We placed the 151 out of 154 targets with a Gaia DR2 parallax in a
CMD (Fig. \ref{fig:CMDoverview}).

Our variability classification had the main goal to start compiling a large
unbiased sample of O and B stars for future asteroseismology. Stars with
multiple individual frequencies of identified coherent standing modes and/or
low-frequency stochastically excited IGWs are suitable of asteroseismic
probing. We found 15 single and 8 binary O and B coherent pulsators (i.e., 15\%
of the monitored stars) and 25 stars with stochastic low-frequency signatures
(16\%). All of these 25 stars are located in the LMC and cover a higher
  mass regime than the galactic targets.  Unlike coherent modes, IGW driven by
  the convective core are independent of an iron opacity bump and should thus
  also be excited in low-metallicity stars. However, the low-frequency signal
  could also be produced by subsurface convection, which is strongly 
metal dependent
  \citep{Cantiello2009}. For all of the variable OB stars, an extension of the
TESS light curve monitoring is needed to provide the frequency precision
required for detailed asteroseismic modeling.  Our initial study reveals the
major potential of the combined long-term (+1 year) TESS, Gaia, and
spectroscopic all-sky monitoring as already outlined by \citet{Kollmeier2017}.


\acknowledgments {\it Acknowledgements:} This paper includes data collected by
the TESS mission. Funding for the TESS mission is provided by the NASA Explorer
Program. Funding for the TESS Asteroseismic Science Operations Centre is
provided by the Danish National Research Foundation (Grant agreement no.:
DNRF106), ESA PRODEX (PEA 4000119301) and Stellar Astrophysics Centre (SAC) at
Aarhus University. We thank the TESS and TASC/TASOC teams for their support of
the present work. This research has made use of the SIMBAD database, operated at
CDS, Strasbourg, France. Some of the data presented in this paper were obtained
from the Mikulski Archive for Space Telescopes (MAST). STScI is operated by the
Association of Universities for Research in Astronomy, Inc., under NASA contract
NAS5-2655. We acknowledge the ESO Science Archive Facility.  The research
leading to these results received funding and/or support from: the European
Research Council (ERC) under the European Union's Horizon 2020 research and
innovation program (grant agreement No. 670519: MAMSIE), the Research
  Foundation Flanders (FWO) under the grant agreement G0H5416N 
(ERC runner-up
  grant), the STFC consolidated grant ST/R000603/1, the Polish NCN grant
  (no.\ 2015/17/B/ST9/02082, 2015/18/A/ST9/00578 
and 2016/21/B/ST9/01126), and
the Natural Science and Engineering Research Council (NSERC) of Canada.
The Flatiron Institute is supported by the Simons Foundation.




\clearpage 

\appendix

\section{Variability classification}
\label{appendix: table}

In Table~\ref{table: stars}, we provide the TESS Input Catalogue (TIC) and Gaia
identification numbers, as well as Gaia magnitudes and colours, \texttt{SIMBAD} spectral types and the classification of the dominant source(s) of variability for each of the 154 O and B stars observed by TESS in sectors 1 and 2 at a 2-min cadence. 

\startlongtable
\begin{longrotatetable}
\begin{deluxetable*}{r c c c c c c c c}
\tablecaption{\label{table: stars}Identification numbers, parameters and
  variability classification of the 154 O and B stars.}
\tablehead{\colhead{TIC} & \colhead{Name} & \colhead{Sp.\,Type} & \colhead{Gaia DR2 ID} & \colhead{$M_G$} & \colhead{BP -- RP}  & \colhead{\# Spectra} & \colhead{Instrument} & \colhead{Var.\,Type} \\
& & \colhead{\texttt{SIMBAD}}	&	&	\colhead{(mag)}	&	\colhead{(mag)}}
\startdata
12359289			&	HD~225119	&	Apsi	&	2333119869770412288   &   -0.98   &   -0.12	&	1	&	F	&	rot	\\
\textit{29990592}			&	HD~268623	&	B2Ia	&	4657589254497529344   &   -4.10   &   0.07	&	1	&	F	&	SLF\\
\textit{30110048}			&	HD~268653	&	B3Ia	&	4765410903770283008   &   -2.02   &   -0.22	&	1	&	F	&	SLF\\
\textit{30268695}		&	HD~268809	&	B1Ia	&	5290767631226220032   &   -2.20   &   -0.10	&	1	&	F	&	SLF\\
\textit{30275662}			&	Sk-66~27	&	B3Ia	&	4661771801060664064   &   -3.37   &   -0.04	&	4	&	U	&	SLF\\
\textit{30275861}			&	[HT83] alf	&	O6V		&	4662155839823561216	&	1.47	&	-0.13	&	-	&	-	&	rot\\
\textit{30312676}			&	HD~268726	&	B2Ia	&	4662153812635167488   &   -2.87   &   0.01	&	2	&	U	&	SLF\\
\textit{30312711}			&	BI 42		&	O8V		&	-	&	-	&	-	&	-	&	-	&	SLF / rot / EB?\\
\textit{30317301}			&	HD~268798	&	B2Ia	&	4661511422956353152   &   -3.02   &   0.00	&	-	&	-	&	EV / rot / SLF	\\
\textit{30933383}			&	Sk-68~39	&	B2.5Ia	&	4662153469001983104   &   -2.97   &   0.12	&	2	&	F	&	SLF\\
\textit{31105740}			&	TYC~9161-925-1	&	B0.5Ia	&	4655458160445111552   &   -2.93   &   -0.15	&	1/3	&	F/U	&	SLF	\\
\textit{31181554}			&	HD~269050	&	B0Ia	&	4661392533937464448   &   -2.05   &   -0.22	&	22/12	&	X/U	&	SLF	\\
31867144			&	HD~22252	&	B8IV	&	4671120982756449664   &   -1.11   &   -0.01	&	-	&	-	&		rot / SPB + \bcep hybrid?	\\
33945685			&	HD~223118	&	B9.5V	&	2338752697903817216   &   1.29   &   0.04	&	-	&	-	&	instr? ($\nu_{\rm inst} \simeq 2.8$~d$^{-1}$)/puls	\\
38602305			&	HD~27657~AB	&	B9IV	&	4676067719930656640   &   0.28   &   -0.11	&	-	&	-	&	rot	\\
\textit{40343782}			&	HD~269101	&	B3Iab	&	5288240197589081728   &   -3.18   &   0.04	&	-	&	-	&	SLF / SPB?	\\
41331819			&	HD~43107	&	B8V	&	5282761464287879296   &   0.29   &   -0.09	&	84/12	&	X/U	&	rot / outburst?	\\
47296054			&	HD~214748	&	B8Ve	&	6622561673163632768   &   -1.38   &   -0.10	&	26	&	U	&	rot / SPB / Be	\\
49687057		&	HD~220787	&	B3III	&	4657674745869490304   &   -2.38   &   0.15	&	6/18	&	F/X	&	instr / binary?	\\
53992511			&	HD~209522	&	B4IVe	&	2436569757731466112   &   -2.04   &   -0.25	&	3/18/1/15	&	F/X/E/U	&	rot / SPB / Be	\\
55295028			&	HD~33599	&	B2Vpe	&	6619440159652409728   &   -1.60   &   -0.23	&	1	&	F	&	rot / SPB / Be	\\
66497441			&	HD~222847	&	B9V	&	2391220091406075648   &   -0.10   &   -0.08	&	1/42	&	F/U	&	EV / rot? / SPB?	\\
69925250			&	V* HN Aqr	&	B	&	2402031280004432512   &   -1.58   &   -0.30	&	4	&	U	&	known \bcep	\\
89545031			&	HD~223640	&	A0VpSiSr	&	2390144081839340288   &   0.10   &   -0.15	&	33	&	U	&	rot	\\
92136299			&	HD~222661~A	&	B9V	&	2419885149815948416   &   0.99   &   -0.06	&	28	&	U	&	SPB+Be-type mini-outbursts / rot\\
115177591		&	HD~201108	&	B8IV/V	&	6775980889978989824   &   -0.16   &   -0.07	&	-	&	-	&	SPB	\\
139468902		&	HD~213155	&	B9.5V	&	6521195703338406272   &   1.01   &   -0.05	&	1/63/1	&	F/X/U	&	rot / SPB?	\\
141281495		&	HD~37854	&	B9/9.5V	&	4648666996817764736   &   0.57   &   -0.02	&	-	&	-	&	rot?	\\
\textit{149039372}		&	HD~34543	&	B8V	&	4663645952980782848   &   0.30   &   -0.05	&	-	&	-	&	rot? / SPB?\\
149971754		&	HD~41297	&	B8Ib	&	5482011113182765696   &   -0.26   &   -0.11	&	-	&	-	&	SPB	\\
150357404		&	HD~45796	&	B6V	&	5477233430220256384   &   -0.44   &   -0.17	&	1	&	F	&		SPB	\\
150442264		&	HD~46792	&	B2V	&	4760693797025929344   &   -3.24   &   -0.21	&	-	&	-	&	EB	+ puls/rot\\
152283270		&	HD~208433	&	B9.5V	&	6588214059487740672   &   0.25   &   -0.01	&	-	&	-	&	instr / binary(transit)	\\
167045028		&	HD~45527	&	B9IV	&	5279546835885790720   &   0.01   &   -0.03	&	-	&	-	&	rot	\\
167415960		&	HD~48467	&	B8/9V	&	5266733784509615616   &   0.11   &   -0.08	&	1	&	F	&	const?	\\
167523976		&	HD~49193	&	B2V	&	4660166788974686976   &   -1.48   &   -0.22	&	-	&	-	&	SPB	\\
176935619		&	HD~49306	&	B9.5/A0V	&	5280155179351701504   &   1.02   &   -0.05	&	-	&	-	&	instr? ($\nu_{\rm inst} \simeq 2.8$~d$^{-1}$)	\\
176955379		&	HD~49531	&	B8/9Vn	&	5279020208473101056   &   0.45   &   -0.04	&	-	&	-	&	SPB / rot	\\
177075997		&	HD~51557	&	B7III	&	5266581089830743296   &   -0.99   &   -0.15	&	1	&	F	&	instr / rot?	\\
\textit{179308923}		&	HD~269382	&	O9.5Ib	&	4657651857940816640   &   -2.57   &   -0.05	&	-	&	-	&	SLF? / SPB?\\
\textit{179574710}		&	HD~271213	&	B1Iak	&	4661778810447390720   &   -1.47   &   -0.09	&	-	&	-	&	 SLF\\
\textit{179637387}		&	[OM95] LH 47-373A	&	B1Ia	&	4651354886129065600   &   -1.52   &   -0.10	&	-	&	-	&	SLF	\\
\textit{179639066}		&	HD~269440	&	B1Ib	&	4658741370886084992   &   -2.17   &   -0.10	&	2	&	F	&	SLF	\\
182909257		&	HD~6783	&	Ap Si	&	4635279171434162304   &   0.65   &   -0.04	&	-	&	-	&	rot	\\
197641601		&	HD~207971	&	B8IV-Vs	&	6586825380598277248   &   -1.97   &   2.13	&	22	&	U	&	instr / rot	\\
206362352	&	HD~223145	&	B3V	&	5283754052709233920   &   -2.30   &   -0.22	&	1/3/302	&	F/X/U	&	SPB	\\
206547467		&	HD~210780	&	B9.5/A0	&	6819470079550296960   &   1.74   &   0.07	&	-	&	-	&	rot / const? / SPB?	\\
207176480		&	HD~19818	&	B9/A0Vn(e)	&	4723685987980665088   &   1.59   &   0.25	&	3	&	F	&	SPB? / SLF? / Be	\\
207235278		&	HD~20784	&	B9.5V	&	4733510055655723392   &   -0.98   &   0.01	&	-	&	-	&	EV / rot	\\
220430912		&	HD~31407	&	B2/3V	&	6522301330997312128   &   -1.06   &   -0.25	&	84	&	X	&	EB + puls	\\
224244458		&	HD~221507	&	B9.5IIIpHgMnSi	&	6538585991555664128   &   0.59   &   -0.12	&	3/30	&	F/U	&	rot	+ mini-outburst?\\
229013861		&	HD~208674	&	B9.5V	&	6612822091790516480   &   1.07   &   -0.04	&	-	&	-	&	rot	\\
230981971		&	HD~10144	&	B6Vpe	&	-   &  -   &   -	&	217/270	&	F/U	&	rot / SPB? / Be	\\
231122278		&	HD~29994	&	B8/9V	&	4656238611846958208   &   0.53   &   0.04	&	-	&	-	&	rot / SPB?	\\
238194921		&	HD~24579	&	B7III	&	4627113682690040576   &   -0.35   &   -0.04	&	-	&	-	&	rot / SPB	\\
259862349		&	HD~16978	&	B9Va	&	4695167130257150592   &   0.66   &   -0.07	&	6/9/20	&	F/X/U	&	instr	\\
260128701		&	HD~42918	&	B4V	&	5494534348761557888   &   -1.01   &   -0.25	&	-	&	-	&	SPB	\\
260131665		&	HD~42933	&	B1/2(III)n	&	5499415974230271488   &   -3.45   &   -0.36	&	4	&	F	&	EB	+ \bcep\\
260368525		&	HD~44937	&	B9.5V	&	5494983804202264192   &   -0.18   &   0.11	&	-	&	-	&	SPB?	\\
260540898		&	HD~46212	&	B8IV	&	5496276314480471040   &   -0.55   &   -0.02	&	-	&	-	&	rot?	\\
260640910		&	HD~46860~AB	&	B9IVn+A8V:p?	&	5482771807727582080   &   -0.73   &   -0.08	&	2	&	F	&	SPB/Be	\\
260820871		&	HD~218801	&	B9.5V(n)	&	6381543153782126464   &   0.60   &   -0.05	&	-	&	-	&	rot / binary?	\\
261205462		&	HD~40953	&	B9V	&	4623532264081294464   &   0.66   &   -0.08	&	1	&	F	&	SPB / rot 	\\
262815962		&	HD~218976~AB	&	B9.5/A0V	&	6500025053617700992   &   1.32   &   -0.01	&	-	&	-	&	rot / SPB?	\\
270070443		&	HD~198174	&	B8II	&	6805364208656989696   &   -0.31   &   -0.08	&	-	&	-	&	rot	\\
270219259		&	HD~209014~AB	&	B8/9V+B8/9	&	6617682865193265536   &   -1.40   &   0.23	&	3	&	F	&	instr / \bcep + outburst		\\
270557257		&	HD~49835	&	B9.5V	&	5211969859107211136   &   1.40   &   0.14	&	-	&	-	&	instr ($\nu_{\rm inst} \simeq2.8$~d$^{-1}$)	\\
270622440		&	HD~224112	&	B8V	&	2314214110928211712   &   -0.08   &   -0.10	&	3	&	F	&	EB (contamination!)	\\
270622446		&	HD~224113	&	B5/8	&	2314213698611350144   &   -0.78   &   -0.12	&	36/18	&	F/X	&	EB	\\
271503441		&	HD~2884~AB	&	B8/A0	&	4900927434176620160   &   1.12   &   -0.08	&	11	&	F	&	SPB? / outburst?	\\
271971626		&	HD~62153~AB	&	B9IV	&	5214590201474858624   &   -0.01   &   -0.01	&	-	&	-	&	rot\\
\textit{276864600}		&	HD~269777	&	B3Ia	&	4661270350708775296   &   -1.90   &   -0.05	&	5	&	F	&	SLF	\\
\textit{277022505}		&	HD~269786	&	B1I	&	4657274283151403520   &   -3.66   &   -0.00	&	-	&	-	&	SLF	\\
\textit{277022967}		&	HD~37836	&	B0e(q)	&	4657280639705552768   &   -4.54   &   0.34	&	12/70	&	F/X	&	rot / SPB?/ SLF?/ Be	\\
\textit{277099925}		&	HD~269845	&	B2.5Ia	&	4661439439319405184   &   -2.67   &   -0.04	&	-	&	-	&	SLF	\\
\textit{277103567}		&	HD~37935	&	B9.5V	&	4660284883361750912   &   -0.48   &   -0.04	&	2/30	&	F/X	&	rot / EV?	\\
\textit{277172980}		&	HD~37974	&	B0.5e	&	4657658356271368064   &   -4.36   &   0.51	&	11/6	&	F/X	&	SLF? / Be	\\
\textit{277173650}		&	HD~269859	&	B1Ia	&	4658882353222625920   &   -3.23   &   -0.16	&	40	&	X	&	SLF / instr?	\\
\textit{277298891}		&	Sk-69~237	&	B1Ia	&	4657679659311713024   &   -1.39   &   0.39	&	1	&	F	&	SLF	\\
277982164		&	HD~54239	&	B9.5/A0III/IV	&	5211241295215037696   &   -0.01   &   0.16	&	36	&	U	&	rot / SPB	\\
278683664		&	HD~47770	&	B9.5V	&	5484286625513030016   &   0.76   &   0.04	&	-	&	-	&	const??	\\
278865766		&	HD~48971	&	B9V	&	5496814662861807360   &   0.46   &   -0.02	&	-	&	-	&	const??\\
278867172		&	HD~49111	&	B9.5V	&	5497973449334379904   &   0.96   &   -0.04	&	-	&	-	&	const??	\\
279430029		&	HD~53048	&	B5/7Vn(e)	&	5484134029618653952   &   -1.97   &   -0.02	&	-	&	-	&	rot / Be	\\
279511712		&	HD~53921~AB	&	B9III+B8V	&	5480486644608749696   &   -0.12   &   -0.17	&	-	&	-	&	rot	\\
279957111	&	HD~269582	&	WN10h	&	4658481718680657792   &   -3.51   &   0.33	&	211	&	X	&	rot	\\
280051467		&	HD~19400~AB	&	B8III/IV	&	4645479443883933824   &   -0.45   &   -0.16	&	1/238	&	F/U	&	rot	\\
280684074		&	HD~215573	&	B6V	&	6351882320090933248   &   -0.70   &   -0.19	&	1/8	&	F/U	&	SPB	\\
281703963		&	HD~4150~A	&	A0IV	&	4908022136034353152   &   -0.12   &   0.02	&	1	&	F	&	hybrid SPB/$\delta$ Sct	\\
281741629		&	CD-56~152	&	Be	&	4908668373993964032   &   -3.05   &   -0.23	&	2/6	&	F/U	&	rot / Be	\\
293268667		&	HD~47478	&	B9V	&	5477090356269215616   &   0.60   &   -0.08	&	-	&	-	&	SPB/ rot?	\\
293973218		&	HD~54967	&	B4V	&	5478303942228108288   &   -2.08   &   -0.21	&	-	&	-	&	rot / SPB	\\
294747615		&	HD~30612 	&	B8II/III(pSi)	&	4654833539071572736   &   -0.33   &   -0.15	&	1	&	F	&	rot / SLF?	\\
294872353		&	HD~270754	&	B1.5Ia	&	4657655435693905408   &   -4.36   &   -0.02	&	1/20	&	F/U	&	SLF?	\\
300010961		&	HD~55478	&	B8III	&	5280667689208638976   &   -0.00   &   -0.14	&	-	&	-	&	rot / \bcep?\\
300325379		&	HD~58916	&	B1.5Ia	&	5281254278662916736   &   0.41   &   0.01	&	-	&	-	&	rot?	\\
300329728		&	HD~59426	&	B9V	&	5267524573886888320   &   0.72   &   0.05	&	-	&	-	&	rot	\\
300744369		&	HD~63928	&	B9V	&	5270696836731782144   &   1.08   &   0.01	&	-	&	-	&	rot	\\
300865934		&	HD~64484	&	B9V	&	5275049837626917504   &   0.04   &   -0.06	&	-	&	-	&	instr / outburst?	\\
306672432		&	HD~67252	&	B8/9V	&	5271283391825477248   &   0.76   &   -0.01	&	-	&	-	&	const? / rot?	\\
306824672		&	HD~68221	&	B9V	&	5270992635422817792   &   0.05   &   0.04	&	-	&	-	&	rot? / SPB?	\\
306829961		&	HD~68520~AB	&	B5III	&	5270986008289935232   &   -2.44   &   -0.15	&	-	&	-	&	SPB	\\
307291308		&	HD~71066	&	B9/A0IV	&	5221286296008492416   &   0.06   &   -0.13	&	13	&	U	&	instr / rot?	\\
307291318		&	HD~71046~AB	&	B9III/IV	&	5221286158569529344   &   -0.17   &   -0.09	&	-	&	-	&	instr / rot?	\\
307993483		&	HD~73990	&	B7/8V	&	5221605085661325056   &   0.40   &   -0.07	&	-	&	-	&	\bcep? / SPB	\\
308395911		&	HD~66591	&	B4V	&	5479669466951012224   &   -1.33   &   -0.15	&	2	&	F	&	SPB	\\
308454245		&	HD~67420	&	B9V	&	5275660856854591360   &   0.39   &   0.07	&	-	&	-	&$\delta$ Sct	\\
308456810		&	HD~67170	&	B8III/IV	&	5289613208437434240   &   0.05   &   0.04	&	-	&	-	&	rot?\\
308537791		&	HD~67277	&	B8III	&	5290024533163062144   &   0.18   &   -0.01	&	-	&	-	&	rot	\\
308748912		&	HD~68423	&	B6V	&	5277219758184463488   &   -0.74   &   -0.06	&	-	&	-	&	SLF? / outburst? / instr	\\
\textit{309702035}		&	HD~271163	&	B3Ia	&	4660142634076536320   &   -3.61   &   -0.03	&	-	&	-	&	SLF	\\
313934087		&	HD~224990	&	B3/5V	&	2320885329010329216   &   -1.32   &   -0.19	&	4/18	&	F/X	&	SPB / rot	\\
327856894		&	HD~225253	&	B8IV/V	&	4701860922688030720   &   -0.80   &   -0.14	&	3/109	&	F/U	&	rot? / outburst?/ instr?	\\
349829477		&	HD~61267	&	B9/A0IV	&	5292390815325246976   &   1.07   &   0.09	&	-	&	-	&	const?	\\
349907707		&	HD~61644	&	B5/6IV	&	5289291395127135360   &   -0.35   &   0.23	&	-	&	-	&	EB	+ puls\\
350146577		&	HD~63204	&	ApSi	&	5288156497263051136   &   0.65   &   -0.03	&	-	&	-	&	rot	\\
350823719		&	HD~41037	&	B3V	&	5275962500997316864   &   -0.88   &   -0.23	&	5/6	&	F/U	&	SPB? / rot \\
354671857		&	HD~14228	&	B8IV	&	4936685751335824896   &   0.15   &   3.16	&	10/79	&	F/U	&	SPB? / rot	\\
355141264		&	HD~208495	&	B9.5V	&	6561093750492347136   &   1.30   &   -0.03	&	-	&	-	&	const?\\
355477670		&	HD~220802	&	B9V	&	6525840590207613824   &   0.34   &   -0.09	&	8	&	F	&	const?\\
355653322		&	HD~224686	&	B8V	&	6485326438580933888   &   -0.74   &   -0.07	&	52	&	F	&	rot? / outburst? / instr?	\\
358232450		&	HD~6882~A	&	B6V+B9V	&	4913847589156808960   &   0.10   &   -0.18	&	24/33	&	F/U	&	EB	\\
358466708		&	CD-60~1931	&	B7	&	5290739387520374912   &   -0.04   &   -0.01	&	-	&	-	&	rot / SPB	\\
358467049		&	CPD-60~944 AB	&	B8pSi	&	5290722929205920640   &   0.18   &   0.06	&	-	&	-	&	rot	\\
358467087		&	CD-60~1929	&	B8.5IV	&	5290722860486442752   &   0.36   &   0.13	&	-	&	-	&	SPB / EV? / rot? /SLF?/ instr?\\
364323837		&	HD~40031	&	B6III	&	4758153203612698624   &   -1.44   &   -0.13	&	4	&	F	&	rot? / SPB?\\
364398190		&	CD-60~1978	&	B8.5IV-V	&	5290816009737675392   &   0.64   &   0.05	&	-	&	-	&	rot? / SLF?	\\
364398342		&	HD~66194	&	B3Vn	&	4776318613169946624   &   -2.36   &   -0.33	&	97	&	U	&	rot? / SPB? / Be	\\
364421326		&	HD~66109	&	B9.5V	&	5287999576341359360   &   -0.18   &   0.08	&	-	&	-	&	rot?\\
369457005		&	HD~197630	&	B8/9V	&	6681797793393053696   &   0.44   &   -0.11	&	1/15	&	F/U	&	rot / SPB?\\
370038084		&	HD~26109	&	B9.5/A0V	&	4666380781970681472   &   1.65   &   0.19	&	-	&	-	&	const?\\
372913233		&	HD~65950	&	B8III	&	5290671733195996416   &   -1.20   &   0.06	&	5	&	F	&	const? / outburst? / instr?\\
372913582		&	CD-60~1954	&	B9.5V	&	5290725643625189504   &   0.29   &   0.11	&	-	&	-	&	const? / outburst? / instr?	\\
372913684		&	HD~65987	&	B9.5IVpSi	&	5290820682661822848   &   -0.67   &   -0.02	&	-	&	-	&	rot	\\
\textit{373843852}		&	HD~269525	&	B0I	&	4658678973606061696   &   -2.39   &   -0.32	&	-	&	-	&	 SPB? / rot?\\
\textit{389921913}		&	HD~270196	&	B1.5Ia	&	4662413606586588160   &   -2.83   &   0.09	&	1/8	&	F/U	&	SLF	\\
\textit{391810734}		&	HD~269655	&	B0Ia	&	4661289630893455488   &   -2.51   &   -0.11	&	2	&	F	&	SLF	\\
\textit{391887875}		&	HD~269660	&	B2Ia	&	4657362548954664192   &   -3.36   &   -0.14	&	-	&	-	&	SLF	\\
\textit{404768847}		&	VFTS~533	&	B0Ia	&	4651835308326981504   &   -1.95   &   -0.06	&	-	&	-	&	SPB? / SLF?	\\
\textit{404768956}		&	Cl* NGC 2070 Mel 12	&	B0.5Ia	&	4658479691454645504   &   -2.91   &   0.03	&	-	&	-	&	SLF? / SPB? / instr?	\\
\textit{404796860}		&	HD~269920	&	B3Ia	&	4657652476416109184   &   -3.84   &   0.04	&	-	&	-	&	SLF	\\
\textit{404852071}		&	Sk-69~265	&	B3I	&	4660612881443486720   &   -2.93   &   -0.09	&	-	&	-	&	SLF	\\
\textit{404933493}		&	HD~269997	&	B2.5Ia	&	4660135826507549824   &   -3.14   &   -0.06	&	1	&	F	&	SLF	\\
\textit{404967301}		&	HD~269992	&	B2.5Ia	&	4657636675267019008   &   -3.51   &   0.18	&	2	&	F	&	SLF	\\
410447919		&	HD~64811	&	B4III	&	-	&	-	&	-	&	-	&	-	&	rot/SPB?	\\
410451677		&	HD~66409	&	B8IV/V	&	5290769211774211968   &   0.32   &   0.05	&	4	&	F	&	rot	\\
419065817		&	HD~1256	&	B6III/IV	&	2364986843479227392   &   0.13   &   -0.13	&	-	&	-	&	rot / SPB?	\\
\textit{425057879}		&	HD~269676	&	O6+O9	&	4651834616802932608   &   -1.79   &   -0.20	&	-	&	-	&	instr? / binary? / rot?	\\
\textit{425081475}		&	HD~269700	&	B1.5Iaeq	&	4657685534828270976   &   -2.78   &   0.33	&	1/48	&	F/X	&	SLF / rot? / Be	\\
\textit{425083410}		&	HD~269698	&	O4Ia	&	4660121743354291328   &   -2.38   &   -0.34	&	6	&	U	&	rot / SPB? / SLF?	\\
\textit{425084841}		&	TYC~8891-3638-1	&	B1Ia	&	4658474743652257664   &   -3.93   &   0.04	&	2/6	&	F/U	&	SLF	\\
441182258		&	HD~210934	&	B7V	&	6618669608159645312   &   -0.53   &   -0.14	&	-	&	-	&	instr / SPB/outburst?	\\
441196602		&	HD~211993	&	B8/9V	&	6615398767225726848   &   0.46   &   -0.09	&	2/9	&	U/X	&	const?	\\
469906369		&	HD~212581~AB	&	B9Vn+G0V	&	6404338508023617664   &   -0.10   &   -0.03	&	19	&	U	&	SPB / \bcep / instrumental?	\\
\enddata
\tablecomments{EB = eclipsing binary, EV = ellipsoidal variable, rot = Rotational modulation, SPB, \bcep, SLF = stochastic low-frequency signal, instr = instrumental, const = constant, puls = pulsational signal not clearly identified in any of the previous categories, and $\delta$ Sct = $\delta$ Scuti star. We provide and overview of which stars have high-resolution spectra available in the ESO archive. U = UVES, F = FEROS, X = X-SHOOTER, E = ESPRESSO. Stars with TIC numbers in \textit{italics} are LMC members.}
\end{deluxetable*}
\end{longrotatetable}


\section{TESS light curves and amplitude spectra}
\label{appendix: figures}

The light curves and amplitude spectra for the 154 O and B stars considered in this work.

\begin{figure*}
\figurenum{4}
\plotone{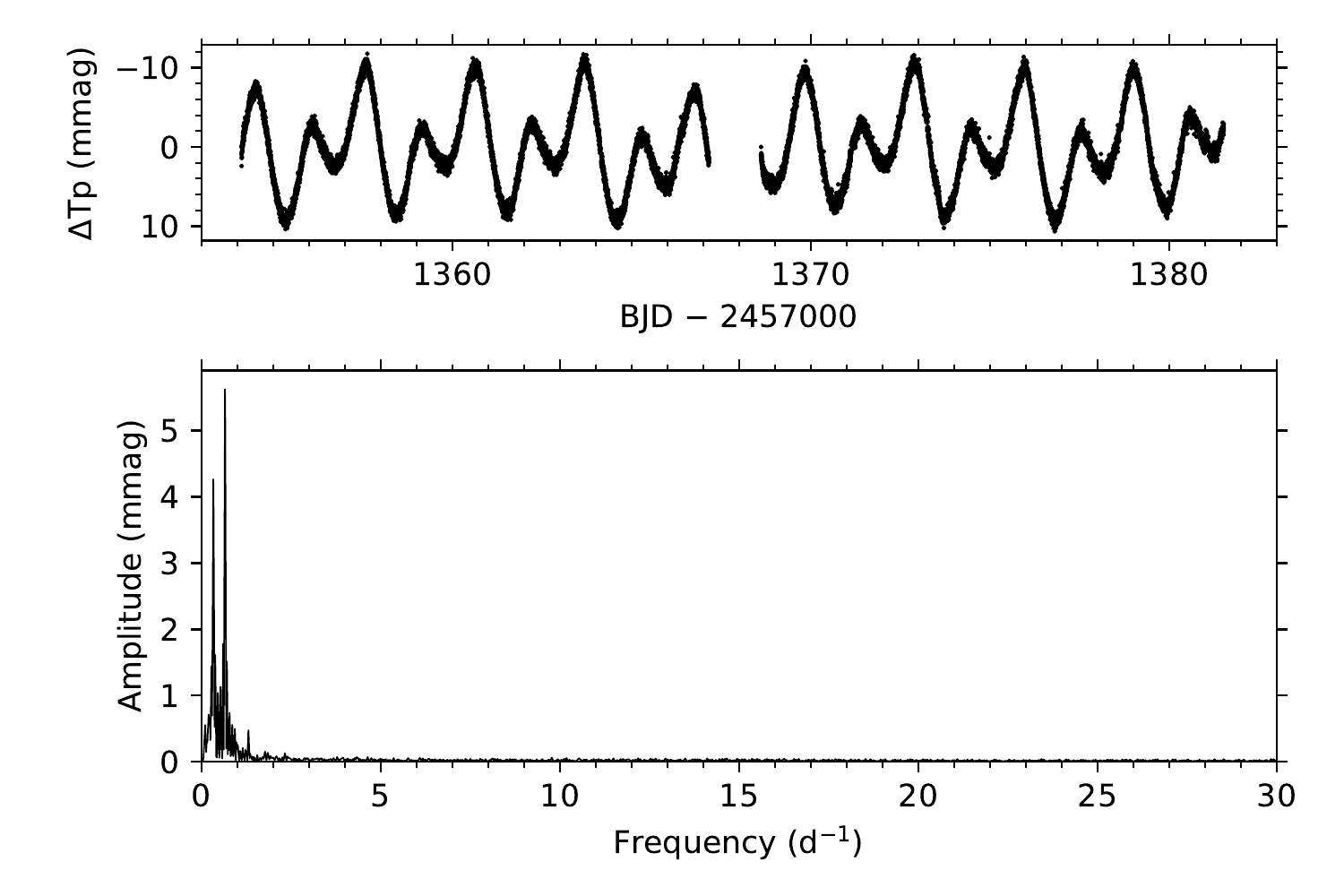}
\caption{The TESS light curve (top) and amplitude spectrum (bottom) of TIC~12359289. The complete figure set (154 images) for all stars considered in this work is available in the online journal.}
\end{figure*}


\end{document}